\documentclass[%
 reprint,
nofootinbib,
 amsmath,amssymb,
 aps,
floatfix,
]{revtex4-2}

\usepackage[normalem]{ulem}
\usepackage{graphicx}% Include figure files
\usepackage{dcolumn}% Align table columns on decimal point
\usepackage{bm}% bold math
\usepackage{ulem}%for strike-through
\usepackage{tikz}
\usetikzlibrary{shapes}
\usepackage{lipsum}
\usepackage{comment}
\usepackage{xcolor}
\usepackage{multirow}
\usepackage{lipsum}
\usepackage{appendix}
\newcommand{\bhexagon}{\mathord{\raisebox{0.75pt}{\tikz{\node[draw,scale=.70,regular polygon, regular polygon sides=6,fill=red](){};}}}} %hexagono red
\newcommand{\bbhexagon}{\mathord{\raisebox{0.75pt}{\tikz{\node[draw,scale=.70,regular polygon, regular polygon sides=6,fill=blue](){};}}}} %hexagono blue
\newcommand{\blackhexagon}{\mathord{\raisebox{0.75pt}{\tikz{\node[draw,scale=.70,regular polygon, regular polygon sides=6,fill=black](){};}}}} %hexagono blue
\newcommand{\QuadAndre}{\mathord{\raisebox{0.75pt}{\tikz{\node[draw,scale=.70,regular polygon, regular polygon sides=4,fill=yellow](){};}}}} %quadrado
\newcommand{\SCAndre}{\mathord{\raisebox{0.75pt}{\tikz{\node[draw, shape = semicircle, star points= 6, scale=.70, fill=blue](){};}}}} %Estrela

\begin{document}

%\preprint{APS/123-QED}

\title{A schematic model for the direct cross-section in reactions induced by exotic and stable projectiles}
%\thanks{A footnote to the article title}
\author{A.~Serra}
\email{aserra@if.usp.br}
\affiliation{Departamento de Física Nuclear, Instituto de Física da Universidade de São Paulo - IFUSP, São Paulo, SP, Brasil.}
\author{R.~Lichtenth\"aler}
 \email{rubens@if.usp.br}
\affiliation{Departamento de Física Nuclear, Instituto de Física da Universidade de São Paulo - IFUSP, São Paulo, SP, Brasil.}
 \author{O.~C.~B. Santos}
 \affiliation{Departamento de Física Nuclear, Instituto de Física da Universidade de São Paulo - IFUSP, São Paulo, SP, Brasil.}
%\email{osvaldo.santos@usp.br}
\author{K.~C.~C. Pires}
\affiliation{Departamento de Física Nuclear, Instituto de Física da Universidade de São Paulo - IFUSP, São Paulo, SP, Brasil.}
%\email{kelly@if.usp.br}
\author{U. Umbelino}
\affiliation{Departamento de Física Nuclear, Instituto de Física da Universidade de São Paulo - IFUSP, São Paulo, SP, Brasil.}

%\affil{IFUSP - Instituto de Física da Universidade de São Paulo.}

%\date{\today}% It is always \today, today,
             %  but any date may be explicitly specified

\begin{abstract}
A geometric model for the direct contribution of the reaction cross section induced by light ions on different targets is presented. The model separates the total reaction cross section into two components, one for total fusion and another for direct reactions. We show that the direct part scales as $2 \pi Ra$, where $R$ is related to the nuclear radius and $a$ is the width of a ring, which is related to the nuclear diffuseness. A simple expression is presented to calculate the radius $R$ and the width parameter $a$ in terms of the masses and charges of the system. The method is applied to experimental data of exotic, weakly bound, and strongly bound  projectiles on several targets. Different diffuseness parameters were obtained for different types of projectiles: exotic \textit{n}-rich, stable weakly bound, stable strongly bound and exotic \textit{p}-rich exotic projectiles.
\end{abstract}

\maketitle

%\tableofcontents

%-------------------------------------------------
\section{Introduction}
\label{sec:intro}
%------------------------------------------------

Low energy reactions induced by light weakly bound and exotic projectiles have been extensively studied in the last two decades \cite{benjamim,kelly,kelly1,kelly2,uiran2019,Keeley,pakou,pakou1,6Li59Co,osvaldo}.
A larger total reaction cross section has been observed in systems involving exotic \textit{n}-rich projectiles, in comparison with stable weakly bound and strongly bound projectiles.  Before comparing cross sections of systems with different masses, it is necessary to re-scale the cross sections to remove trivial geometric effects such as different radii and Coulomb barriers. Different methods have been proposed to reduce the total reaction cross sections
 \cite{CANTO2009,CANTO20151,gomes1,gomes2,kelly2,pfaria,Disentangling_Canto,Mohr_alfa}. Application of these methods to experimental data shows that there are three main classes of reduced cross sections. Exotic neutron rich projectiles such as $^6$He usually present the larger reduced cross section followed by the weakly bound such as $^{6,7,8}$Li, $^7$Be and, finally, the strongly bound projectiles such as alpha particles, $^{12}$C and $^{16}$O. Although this enhancement has been observed mainly in reactions with heavy targets, it has been reported on light targets as well \cite{kelly1}. The reasons for this enhancement are not yet completely understood. Light nuclei usually exhibit a strong cluster structure which is expected to play an important role in the reaction mechanisms. In  particular, at low energies, around and  below the Coulomb barrier, coupled channels effects are expected to be more important,  however, it is not clear how fusion and direct cross sections are affected by coupled channels effects.   

In collisions induced by projectiles with alpha-structure, a large yield of alpha particles has been observed in the spectra as early as 1961 \cite{alfa_yield_antigo} and, latter, in other nuclides such as $^6$He, $^{7}$Li as well \cite{ribras3}. Investigation of the angular and energy distributions of these fragments indicate that they are produced mainly in direct processes, such as neutron transfer and projectile breakup.  Direct reaction should have quite different characteristics compared to non-direct (compound nucleus) processes, mainly regarding the energy and angular distributions of the reaction products. Due to the very different time scales of direct ($\approx 10^{-22}$ s) and compound nucleus ($\approx 10^{-19}$ s) processes, the angular and energy distributions of the reaction products are expected to be very different. Particles produced by direct processes are expected to have a forward peaked distribution with energies near the energy of the projectile. On the other hand, for processes which occur via compound nucleus formation, a more isotropic angular distribution is expected with an energy distribution shifted toward lower energies. Nevertheless, in practical terms, the experimental separation of direct and non-direct reactions (fusion) is not trivial at low energies. In particular below the Coulomb barrier, the experimental separation between fragments from direct processes and those coming from complete fusion becomes tricky. Reactions such as incomplete fusion can contribute in a region of energies where it is difficult to separate from pure direct processes, requiring the measurement of more involved degrees of freedom such as neutrons and gammas in coincidence. At energies above the barrier the situation improves and it seems to be possible to obtain reliable direct cross sections by measuring only the charged fragments distributions. 

In the present paper, we propose a method to estimate the direct part of the total reaction cross section at energies above the Coulomb barrier. In section \ref{sec:Metodo}, we present the formalism. In  section \ref{sec:Application}, the method is applied to analyse experimental data. In section \ref{sec:Conclusao}, we present the conclusions.

\section{The Method and its application to data.}
\label{sec:Metodo}

At energies above the Coulomb barrier, for weakly deformed nucleii, the total reaction cross section can be calculated using the well know geometric formula, given below:
\begin{equation}
\sigma_R= \pi R_b^2(1-\frac{V_b}{E}) 
\label{eq3}
\end{equation}

\noindent where $R_b$ and $V_b$ are, respectively, the Coulomb radius and the Coulomb barrier. This equation has been used to reduce the total reaction cross section data \cite{eli,kelly2,pfaria, vivi1,vivi2}.  The total reaction cross section tends to $\pi R_b^2$ for $E \gg V_b$ and the term inside parentheses gives its energy dependence as a function of the ratio $E/V_b$. For energies below the barrier, one may use the Wong formula \cite{Wong_Original, WONG_Original_1972} - Eq.[\ref{eq:Wong_Andre}].%which extends the validity of Eq.\ref{eq3} for energies below the barrier given by Eq.\ref{eq:Wong_Andre}.
 %Eq.\ref{eq3} is valid only for energies above the coulomb barrier ($E>V_b$). 

\begin{figure}
    \centering
    \includegraphics[width=0.3\textwidth]{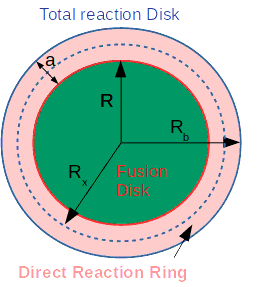}
    \caption{A geometric picture of the model. $R_{b}$ and $R$ determine the total reaction and fusion cross sections disks, respectively. $R_x$ and $a$ stand for radius and width of the direct reaction ring, respectively.}
    \label{fig:ideia}
\end{figure}

\begin{equation}
\sigma_R =  \left[R_b^2\hbar\omega/2E \right] \mathrm{ln}\{1+\mathrm{exp}(2\pi(E-V_b)/\hbar\omega)\}
\label{eq:Wong_Andre}
\end{equation}

It is easy to show that, for energies slightly above the Coulomb barrier, Eq.[\ref{eq:Wong_Andre}] reduces exactly to the geometric Eq.[\ref{eq3}].

The next step is to write the total reaction cross section as the sum of two contributions, one for total fusion and another for direct reactions:
\begin{equation}
    \sigma_R=\sigma_{fus}+\sigma_{dir} 
    \label{eq4}
\end{equation}

Direct reactions are expected to be more peripheral processes, taking place in a limited angular momentum range located near the grazing angular momentum with a certain width. At high energies in \textit{r}-space, this corresponds to a ring of radius $R_x$ and width $a$ which leads to the next relation - see  Fig.\ref{fig:ideia} and Ref.\cite{scaling}: 

\begin{equation}
    \pi R_b^2=\pi R_{fus}^2+2\pi R_xa 
    \label{eq4a}
\end{equation}

\noindent Thus, the simple substitution  $\pi R_b^2 \rightarrow 2\pi R_x a$ in Eqs.[\ref{eq3}, \ref{eq:Wong_Andre}] leads to  the following expression for the total cross section for direct processes:

\begin{equation}
\sigma_{dir}= 2 \pi R_xa\left(1-\frac{V_b}{E}\right) 
\label{eq3a}
\end{equation}
\noindent and similarly for Eq.[\ref{eq:Wong_Andre}]. Indeed, as early as 1947, for high energy deuterons (around 190 MeV), similar expression was used in Ref.\cite{Serber}.  Following this geometric approach, we can define $R_x = R_{fus}+a/2$ and $ a=R_b - R_{fus}$. The parameter $a$ is the principal quantity in our methodology. 

From now on, we apply a model \cite{woods, kelly2} which relates the Coulomb $R_b$ and the nuclear $R$ radii. This model provides a quite natural connection with the geometric picture of Fig.\ref{fig:ideia}. From this model:

\begin{equation}
R_b=R+a_n\mathrm{ln}(X)
\label{eq5a}
\end{equation}

\noindent and, in this last expression:
\begin{equation}
R_{fus} \approx R=r_0(A_p^{1/3}+A_t^{1/3})
\label{eq:Scale}
\end{equation}

\noindent with $r_0 = 1.3$ fm standing for the reduced nuclear radius, and $a_n$ is the diffuseness of the nuclear potential, whose standard value for stable ions is $a_n=0.65$ fm. The parameter $X$ is given by \cite{woods}:

\begin{equation}
X=27.1\frac{[A_p^{1/3}+A_t^{1/3}]^2}{Z_pZ_t}
\label{eq6a}
\end{equation}

The parameters of the above equation have been obtained in Ref.\cite{woods} using a real Woods-Saxon nuclear potential that fits the tail of a double folding potential. It is interesting to note that the parameter $X$ has a dependence on A/Z, which seems to be present in the data, as we will show later. It has been shown that fusion cross sections scale as in Eq.[\ref{eq:Scale}] with normal $r_0$ values around 1.2-1.5 fm \cite{gomes2}. Thus, the simple substitution $R \approx R_{fus}$ in Eq.[\ref{eq5a}] provides a formula for the width $a$ of the direct reaction disk, as given below:
\begin{equation}
a=a_n\mathrm{ln}(X) 
\label{eq7a}
\end{equation}

A universal curve for the reduced direct cross section, $\sigma_{red}$, is obtained simply by dividing Eq.[\ref{eq3a}] by $2 \pi R_xa$:

\begin{equation}
\sigma_{red}=1-\frac{1}{x}
\label{uni1}
\end{equation}
\noindent with $x=E/V_b$. In the case of the Wong formula (Eq.[\ref{eq:Wong_Andre}]), a slightly different reduction holds: $\sigma^{\prime}_{red} = \sigma_{dir}E/(aR_x\hbar\omega)$, which leads to:
\begin{equation}
\sigma^{\prime}_{red} = \mathrm{ln}\{1+\mathrm{exp}(2\pi x^{\prime})\}
\label{uni2}
\end{equation}
with $x^{\prime}=(E-V_b)/\hbar\omega$. Recently, Ref.\cite{7Li124Sn}, some transfer direct reaction channels were normalized using the usual Wong formula with free parameters related to barrier shift and separation energies. Both, the reduced cross sections, $\sigma_{red}$ and $\sigma^{\prime}_{red}$, and the independent variables, $x$ and $x^{\prime}$ are dimensionless quantities and will be used to compare this method with experimental data in the next section. 

\section{Application to data}
\label{sec:Application}

This model was applied to analyze experimental data corresponding to the direct part of total reaction cross section. 
%As mentioned in the introduction, it is not always trivial to separate experimental data from direct and non-direct processes. 
We selected the experimental data by two methods:

\setlength{\tabcolsep}{4pt}
\renewcommand{\arraystretch}{1.2}

\begin{table*}[!h]
\scriptsize
\caption{Data evaluation summary. For exotic \textit{n}-rich there are 28 systems; for WB: 71; SB: 35 and for exotic \textit{p}-rich: 12. See text for: $a$, $V_b$ and $R_x$. The symbol $\downarrow V_b$ means: all system's data are bellow the Coulomb barrier. Data are presented as ordered pairs: ($x$; $\sigma_{dir}$). Methods are divided as: 1 - integration of particle/neutron distribution or $\gamma$-spectroscopy; and, 2 - Difference between reaction and fusion cross sections.}
\label{tab:TodosDados}

\begin{tabular}{lccccccp{8.0cm}}
\hline
\hline
System & Fig.\ref{fig:Final} & Ref. & Method & $V_b$ (MeV) & $R_x$ (fm) & $a$ (fm) & \hspace{4cm} Data ($x$; $\sigma_{dir}$ [mbar])\\ \hline
\multicolumn{8}{c}{exotic \textit{n}-rich; $a_n = 1.32(06)$ fm.}\\
\hline
 \hspace{4pt}$^6$He + $^{120}$Sn & \textcolor{red}{$\bigstar$} & \cite{ribras3} & 1 & 12.73 &  10.61 &  3.67 & \begin{tabular}{p{1.5cm}p{1.5cm}p{1.5cm}p{1.5cm}} 1.30; 940.0 & 1.48; 977.0 & 1.53; 1141.0 & 1.35; 1031.0\end{tabular}\\ 
\begin{tabular}{l} 
\hspace{-0pt}$^6$He + $^{209}$Bi\\
\end{tabular} &\begin{tabular}{l}
\textcolor{red}{$\blacktriangledown$} \\
\end{tabular} & \begin{tabular}{l}
\cite{6He209Bi}\\
\end{tabular} & \begin{tabular}{l}
2\\
\end{tabular} &  \begin{tabular}{l}
19.31\\
\end{tabular} & 
\begin{tabular}{l}
11.74\\
\end{tabular} & \begin{tabular}{l}
3.33\\
\end{tabular} & \begin{tabular}{p{1.5cm}p{1.5cm}} 0.91; 562.6 & 1.12; 783.8 \end{tabular}\\ 
\hspace{1pt} $^6$He + $^{64}$Zn &\textcolor{red}{$\blacktriangleleft$} & \cite{6He64Zn} & 1 & 8.34 & 9.55 & 3.98 & \begin{tabular}{p{1.5cm}p{1.5cm}} 1.10; 300.0 & 1.49; 1224.0 \end{tabular}\\ 
\begin{tabular}{l}
$^8$He + $
^{197}$Au\\
\\
\end{tabular} &
\begin{tabular}{l}
\textcolor{red}{$\blacksquare$}\\
\\
\end{tabular} & \begin{tabular}{l}
\cite{8He197Au}\\
\\
\end{tabular} & \begin{tabular}{l}
1\\
\\
\end{tabular} & \begin{tabular}{l}
18.18\\
\\
\end{tabular} & \begin{tabular}{l}
11.88\\
\\
\end{tabular} & \begin{tabular}{l}
3.43\\
\\
\end{tabular} & \begin{tabular}{p{1.5cm}p{1.5cm}p{1.5cm}p{1.5cm}} 1.53; 1096.0 & 1.36; 1131.0 & 1.20; 997.0 & 1.03; 761.0 \\ 0.96; 666.0 & 0.88; 259.0 &  0.70; 30.0 & \end{tabular}\\

\hspace{4pt}$^{8}$He + $^{208}$Pb & $\bhexagon$ & \cite{8He208Pb} & 1 & 18.68 & 11.84 & 3.08 & \begin{tabular}{p{1.5cm}p{1.5cm}} 0.82; 229.0 & 1.13; 1264.0\end{tabular}\\ 

\hspace{4pt}$^{11}$Be + $^{64}$Zn &\textcolor{red}{$\blacktriangle$} & \cite{11Be64Zn, 11Be64ZnII} & 1 & 16.67 & 9.67 & 3.16 & \begin{tabular}{p{1.5cm}} 1.47; 1500.0 \end{tabular}\\ 
\begin{tabular}{l}
$^{11}$Li + $^{12}$C\\
\\
\end{tabular} &
\begin{tabular}{l}
\textcolor{red}{$ \blacktriangleright$}\\
\\
\end{tabular} & \begin{tabular}{l}
\cite{11Li9Be12C}\\
\\
\end{tabular} & \begin{tabular}{l}
1\\
\\
\end{tabular} & \begin{tabular}{l}
2.73\\
\\
\end{tabular} & \begin{tabular}{l}
8.36\\
\\
\end{tabular} & 
\begin{tabular}{l}
5.00\\
\\
\end{tabular} & \begin{tabular}{p{1.5cm}p{1.5cm}p{1.5cm}p{1.5cm}} 2.76; 1502.9 & 4.54; 1646.6 & 6.39; 1749.7 & \\
10.8; 1808.7 & & & \end{tabular}\\ 
\begin{tabular}{l}
$^{11}$Li + $^{9}$Be\\
\\
\end{tabular} &\begin{tabular}{l}
\textcolor{red}{$ \blacklozenge$}\\
\\
\end{tabular} & \begin{tabular}{l}
\cite{11Li9Be12C}\\
\\
\end{tabular} & \begin{tabular}{l}
1\\
\\
\end{tabular} & \begin{tabular}{l}
1.80\\
\\
\end{tabular} &	\begin{tabular}{l}
8.32\\
\\
\end{tabular} & \begin{tabular}{l} 5.45\\
\\
\end{tabular} & \begin{tabular}{p{1.5cm}p{1.5cm}p{1.5cm}p{1.5cm}} 4.37; 2332.6 & 7.15; 2406.4 & 9.91; 2409.9 \\
22.1; 1774.0 & & & \end{tabular}\\ 
\hspace{4pt}$^{11}$Li + $^{208}$Pb &$\QuadAndre$ &  \cite{11Li208Pb} & 1 & 28.04 & 11.90 & 2.90 & \begin{tabular}{p{1.5cm}p{1.5cm}} 0.82; 5100.0 & 1.06; 6500 \end{tabular}\\ 
\hline
\multicolumn{8}{c}{Weakly bound (WB); $a_n = 0.66^{+0.16}_{-0.12}$ fm.}\\
\hline
\begin{tabular}{l}
$^6$Li + $^{209}$Bi \\
\\
\end{tabular}&\begin{tabular}{l}
\textcolor{blue}{$\bigstar$}\\
\\
\end{tabular}
& \begin{tabular}{l}
\cite{Jin15}\\
\\
\end{tabular}& 
\begin{tabular}{l}
1\\
\\
\end{tabular}& 
\begin{tabular}{l}
29.79\\
\\
\end{tabular}& \begin{tabular}{l}
10.69\\
\\
\end{tabular}& 
\begin{tabular}{l}
1.24\\
\\
\end{tabular}& \begin{tabular}{p{1.5cm}p{1.5cm}p{1.5cm}p{1.5cm}} 1.63; 699.4 & 1.31; 579.9 & 0.78; 20.0 & 1.04; 337.3 \\
0.91; 113.2 & 0.85; 46.4 & 1.24; 551.1 & \end{tabular}\\ 
\begin{tabular}{l}
$^6$Li + $^{208}$Pb\\
\\
\end{tabular}&
\begin{tabular}{l}
\Large{\textcolor{blue}{$\bullet$}}\\
\\
\end{tabular}& 
\begin{tabular}{l}
\cite{6He209Bi}\\
\\
\end{tabular}& 
\begin{tabular}{l}
2\\
\\
\end{tabular}& 
\begin{tabular}{l}
29.44\\
\\
\end{tabular}& 
\begin{tabular}{l}
11.69\\
\\
\end{tabular}& \begin{tabular}{l}
1.24\\
\\
\end{tabular}&
\begin{tabular}{p{1.5cm}p{1.5cm}p{1.5cm}p{1.5cm}} 1.02; 208.7 & 1.08; 309.8 & 1.14; 430.0 & 1.20; 531.1 \\
1.31; 746.1 & & &  \end{tabular}\\  
\begin{tabular}{l}
$^6$Li + $^{90}$Zn\\
\\
\end{tabular} & 
\begin{tabular}{l}
\textcolor{blue}{$\blacksquare$} \\
\\
\end{tabular} & 
\begin{tabular}{l}
\cite{kuma}\\
\\
\end{tabular} & 
\begin{tabular}{l}
1\\
\\
\end{tabular} & \begin{tabular}{l}
16.49\\
\\
\end{tabular} & 
\begin{tabular}{l}
8.91\\
\\
\end{tabular} & 
\begin{tabular}{l} 1.45 \\
\\
\end{tabular} & \begin{tabular}{p{1.5cm}p{1.5cm}p{1.5cm}p{1.5cm}} 1.70; 607.0 & 1.42; 540.1 & 1.19; 442.4 & 1.07; 344.7 \\
0.96; 159.5 & 0.85; 41.2 & & \end{tabular}\\ 
\begin{tabular}{l}
$^6$Li + $^{59}$Co
\end{tabular}&
\begin{tabular}{l}
\textcolor{blue}{$\blacktriangle$}\\
\end{tabular}& 
\begin{tabular}{l}
\cite{6Li59Co}\\
\end{tabular}& \begin{tabular}{l}
1\\
\end{tabular}& \begin{tabular}{l}
11.75\\
\end{tabular}& \begin{tabular}{l}
8.21\\
\end{tabular}& 
\begin{tabular}{l}
1.58\\
\end{tabular}& \begin{tabular}{p{1.5cm}p{1.5cm}p{1.5cm}p{1.5cm}} 1.34; 317.1 & 1.66; 424.4 & 1.97; 457.0 & 2.29; 471.0 \\
\end{tabular}\\ 
\hspace{4pt}$^7$Li + $^{9}$Be &\textcolor{blue}{$\blacktriangledown$}& \cite{verma_interaction_2010} & 2 & 1.89 & 6.37 & 2.26 & \begin{tabular}{p{1.5cm}p{1.5cm}p{1.5cm}} 4.68; 655.5 & 7.13; 701.2 & 8.91; 776.2 \end{tabular}\\ 
\begin{tabular}{l}
$^7$Li + $^{208}$Pb\\
\\
\end{tabular} & \begin{tabular}{l}
\textcolor{blue}{$\blacktriangleleft$}\\
\\
\end{tabular} & \begin{tabular}{l}
\cite{67Li208Pb}\\
\\
\end{tabular} & 
\begin{tabular}{l}
2\\
\\
\end{tabular} & 
\begin{tabular}{l} 
29.10\\
\\
\end{tabular} & \begin{tabular}{l}
10.95\\
\\
\end{tabular} & \begin{tabular}{l}
1.10
\\
\\
\end{tabular}&
\begin{tabular}{p{1.5cm}p{1.5cm}p{1.5cm}p{1.5cm}} 1.99; 556.6 & 1.46; 304.2  & 1.30; 343.0 & 1.16; 161.8 \\
1.10; 174.8 & 1.03;	168.3 & 0.96; 84.1 & \end{tabular}\\  
\begin{tabular}{l}
$^7$Li + $^{124}$Sn \\
\\
\\
\\
\\
\end{tabular}&
\begin{tabular}{l}
\large{\textbf{\textcolor{blue}{$\ast$}}} \\
\\
\\
\\
\\
\end{tabular}& \begin{tabular}{l}
\cite{7Li124Sn}\\
\\
\\
\\
\\
\end{tabular}& \begin{tabular}{l}
1\\
\\
\\
\\
\\
\end{tabular}& 
\begin{tabular}{l}
19.29\\
\\
\\
\\
\\
\end{tabular}
& 
\begin{tabular}{l}
9.68\\
\\
\\
\\
\\
\end{tabular}& \begin{tabular}{l}
1.42\\
\\
\\
\\
\\
\end{tabular}&
\begin{tabular}{p{1.5cm}p{1.5cm}p{1.5cm}p{1.5cm}} 0.83; 28.2 & 0.88; 50.7 &  0.93; 101.4  & 0.98; 163.4 \\
1.03; 242.3 & 1.08; 332.5 & 1.13; 417.0 & 1.18; 473.3 \\ 
1.23; 501.5 & 1.28; 524.1 & 1.32; 524.1 & 1.37; 541.0 \\ 
1.42; 602.9 & 1.47; 614.2 & 1.52; 625.5 & 1.57; 659.3 \\
1.62; 693.1 & 1.67; 743.8 & 1.72; 777.6 & 1.77; 783.3 
\end{tabular}\\
\hspace{4pt}$^7$Be + $^{28}$Si &\textcolor{blue}{$\bbhexagon$}& \cite{7Be28Si} & 1 & 8.87 & 7.37 & 1.63 & \begin{tabular}{p{1.5cm}p{1.5cm}p{1.5cm}} 1.98; 366.0 & 1.80; 335.0 & 1.17; 111.0\end{tabular}\\
\hspace{4pt}$^8$Li + $^{208}$Pb & \textbf{\Large{\textcolor{blue}{$+$}}} & \cite{KOL02} & 1 & 28.79 & 10.94 & 1.27 & \begin{tabular}{p{1.5cm}} 1.16; 374.5 \end{tabular}\\
\hspace{4pt}$^8$Li + $^{209}$Bi &\textbf{\Large{\textcolor{blue}{$\times$}}} & \cite{8Li209Bi} & 1 & 29.13 & 10.95 & 1.27 & \begin{tabular}{p{1.5cm}} 1.28; 434.2 \end{tabular}\\
\hspace{4pt}$^8$Li + $^{58}$Ni &\textcolor{blue}{$\blacklozenge$}& \cite{osvaldo} & 1 & 11.92 & 8.42 & 1.59 & \begin{tabular}{p{1.5cm}p{1.5cm}p{1.5cm}p{1.5cm}} 1.76; 290.7 & 2.21; 280.7 & 1.92; 289.7 & 2.12; 283.7 \end{tabular}\\
\hspace{1pt} $^{9}$Be + $^{9}$Be &\textbf{\textcolor{blue}{$\downarrow V_b  \blacktriangleright$}} & \cite{9Be9Be} & 1 & 2.54 & 6.52 & 2.23 & \begin{tabular}{p{1.5cm}p{1.5cm}} 0.49; 4.0 & 0.68; 32.0 \end{tabular}\\
\begin{tabular}{l}
$^{9}$Be + $^{208}$Pb \\
     \\
\end{tabular}& 
\begin{tabular}{l}
     \textcolor{blue}{$\SCAndre$} \\
     \\
\end{tabular} & 
\begin{tabular}{l}
     \cite{9Be208Pb,9Be208Pb_Fusao, 9Be208Pb_FusaoII} \\
     \\
\end{tabular}
 & 
 \begin{tabular}{l}
     2 \\
     \\
\end{tabular} & 
\begin{tabular}{l}
     38.78 \\
     \\
\end{tabular} &
\begin{tabular}{l}
     11.04\\
     \\
\end{tabular}
 & 
 \begin{tabular}{l}
     1.26\\
     \\
\end{tabular} &
\begin{tabular}{p{1.5cm}p{1.5cm}p{1.5cm}p{1.5cm}} 0.94; 142.5 & 0.99; 251.6 & 1.04; 356.1  & 1.09; 441.1\\
 1.14; 459.0 & 1.19; 748.2 & 1.24; 647.0 & 1.48; 930.0\\
\end{tabular}\\
\hline
\multicolumn{8}{c}{exotic \textit{p}-rich; $a_n = 0.55$ fm - no uncertain evaluation was possible.}\\
\hline
\hspace{4pt}$^8$B + $^{28}$Si & $\blacktriangle$ & \cite{8B28Si} & 1 & 11.17 & 7.16 & 1.23 & \begin{tabular}{p{1.5cm}} 19.62; 557.0 \end{tabular}\\

\begin{tabular}{l}
$^8$B + $^{58}$Ni\\
\end{tabular} & \begin{tabular}{l}
$\blacksquare$\\
\end{tabular} & \begin{tabular}{l}
\cite{8B58Ni}\\
\end{tabular} & \begin{tabular}{l}
2\\
\end{tabular} & \begin{tabular}{l} 
20.81\\
\end{tabular} & \begin{tabular}{l}
8.28\\
\end{tabular} & 
\begin{tabular}{l}
1.30\\
\end{tabular} & \begin{tabular}{p{1.5cm}p{1.5cm}p{1.5cm}p{1.5cm}} 0.73; 188.0 & 0.94; 190.0 & 1.06; 138.0 & 1.12; 308.0 \end{tabular}\\

\hspace{4pt}$^8$B + $^{64}$Zn &$\blacklozenge$ & \cite{8B64Zn} & 1 & 21.98 & 8.30 & 1.01 & \begin{tabular}{p{1.5cm}} 1.56; 315.0 \end{tabular}\\

\hspace{4pt}$^8$B + $^{208}$Pb &$\blacktriangledown$ & \cite{8B208Pb} & 1 & 49.75 & 10.68 & 0.77 & \begin{tabular}{p{1.5cm}} 0.47; 326.0 \end{tabular}\\
	
\begin{tabular}{l}
$^{17}$F + $^{58}$Ni \\
\end{tabular} & 
\begin{tabular}{l}
$\blackhexagon$ \\
\end{tabular} & 
\begin{tabular}{l}
\textcolor{red}{\cite{17F58Ni}}\\
\cite{8B208Pb}
\end{tabular} &
\begin{tabular}{l}
1 \\
\\
\end{tabular} &
\begin{tabular}{l}
36.12 \\
\\
\end{tabular} &
\begin{tabular}{l}
8.78 \\
\\
\end{tabular} &
\begin{tabular}{l}
0.81 \\
\\
\end{tabular} & \begin{tabular}{p{1.5cm}p{1.5cm}p{1.5cm}p{1.5cm}} \textcolor{red}{1.16; 85.20} &\textcolor{red}{1.25; 106.40} & & \\ 0.93; 80.2 & 1.02; 98.4 & 1.19; 151.1 & 1.35; 134.9
\end{tabular} \\

\hspace{4pt}$^{17}$F + $^{208}$Pb & \Large{\textcolor{black}{$\bullet$}} & \cite{17F208Pb} & 1 & 87.02 & 11.31 & 0.52 & \begin{tabular}{p{1.5cm}p{1.5cm}p{1.5cm}} 1.04; 69.3 & 1.27; 124.7 & 1.81; 103.9 \end{tabular} \\

\hline
\multicolumn{8}{c}{Strongly bound (SB); $a_n = 0.68^{+0.16}_{-0.04}$ fm.}\\
\hline
\hspace{4pt}$^{4}$He + $^{64}$Zn &\textcolor{green}{$\blacktriangleright$}& \cite{alfa64ZnMet1} & 1 & 8.63 & 8.16 & 1.80 & \begin{tabular}{p{1.5cm}p{1.5cm}} 1.31; 336.0 & 1.80; 668.0 \end{tabular}\\
\begin{tabular}{l} 
$^{4}$He + $^{209}$Bi \\
\\
\\
\end{tabular} & \begin{tabular}{c} \Large{\textcolor{green}{$\bullet$}}\\
\\
\\
\end{tabular} & \begin{tabular}{c} \cite{4He209Bi_Elastic}\\
\\
\\
\end{tabular} & \begin{tabular}{c} 
2\\
\\
\\
\end{tabular} & \begin{tabular}{c} 
19.86\\
\\
\\
\end{tabular} & \begin{tabular}{c} 
10.53\\
\\
\\
\end{tabular} & \begin{tabular}{c} 
1.51\\
\\
\\
\end{tabular} & \begin{tabular}{p{1.5cm}p{1.5cm}p{1.5cm}p{1.5cm}} 1.11; 116.6 & 1.24; 12.3 & 1.43; 110.4 & 1.75; 122.7 \\
1.93; 355.8 & 2.10;	355.8 & 2.45; 533.7 & 3.46; 950.9 \\
5.14; 1092.0 & & & \end{tabular}\\
\hspace{4pt}$^{12}$C + $^{13}$C &\textcolor{green}{$\blacktriangleleft$}& \cite{12C13CMet2} & 2 & 5.74 & 6.98 & 1.89 & \begin{tabular}{p{1.5cm}p{1.5cm}p{1.5cm}} 2.76; 186.4 & 4.18; 202.79 & 4.67; 355.89 \end{tabular}\\
\begin{tabular}{l}
$^{12}$C + $^{16}$O\\
\\
\end{tabular} & \begin{tabular}{l}
\textcolor{green}{$\blacktriangledown$}\\
\\
\end{tabular} & \begin{tabular}{l}
\cite{12C16O}\\
\\
\end{tabular} & \begin{tabular}{l}
2\\
\\
\end{tabular} & \begin{tabular}{l}
7.66\\
\\
\end{tabular} & \begin{tabular}{l}
7.12\\
\\
\end{tabular} & \begin{tabular}{l}
1.75\\
\\
\end{tabular} & \begin{tabular}{p{1.5cm}p{1.5cm}p{1.5cm}p{1.5cm}} 1.17; 40.4 & 1.30;5.6 & 1.56; 225.5 & 1.58; 225.3 \\
1.92; 360.1 & & & \end{tabular}\\
\begin{tabular}{l}
$^{12}$C + $^{208}$Pb\\
\end{tabular}& \begin{tabular}{l}
\textcolor{green}{$\blacklozenge$} \\
\end{tabular} & \begin{tabular}{l}
\cite{12C208Pb}\\
\end{tabular} & \begin{tabular}{l}
2\\
\end{tabular} & \begin{tabular}{l}
58.34\\
\end{tabular} & \begin{tabular}{l}
11.12\\
\end{tabular} & \begin{tabular}{l}
0.89\\
\end{tabular} &  \begin{tabular}{p{1.5cm}p{1.5cm}p{1.5cm}p{1.5cm}} 0.95; 2.6 & 0.99; 56.6 & 1.02; 136.6 & 1.05; 198.5
\end{tabular}\\
\hspace{4pt}$^{12}$C + $^{209}$Bi &\textcolor{green}{$\blacktriangle$}& \cite{12C209Bi} & 2 & 59.04 & 11.13 & 0.88 & \begin{tabular}{p{1.5cm}p{1.5cm}} 1.12; 16.0 & 1.17; 71.0 \end{tabular}\\
\begin{tabular}{l}
$^{16}$O + $^{58}$Ni\\
\\
\\
\end{tabular} & 
\begin{tabular}{l}
\textcolor{green}{$\blacksquare$} \\
\\
\\
\end{tabular} &
\begin{tabular}{l}
\cite{16ONi}\\
\\
\\
\end{tabular} & \begin{tabular}{l}
2\\
\\
\\
\end{tabular} & 
\begin{tabular}{l}
32.02\\
\\
\\
\end{tabular} & 
\begin{tabular}{l} 
8.85\\
\\
\\
\end{tabular} & \begin{tabular}{l}
1.08\\
\\
\\
\end{tabular} & \begin{tabular}{p{1.5cm}p{1.5cm}p{1.5cm}p{1.5cm}} 2.94; 474.5 & 2.45; 370.2 & 1.96; 265.9 & 1.71; 208.6 \\
1.47; 119.9 & 1.17; 67.8 & 1.13; 46.9  & 1.08; 36.5 \\
1.03; 15.6 & 0.98; 10.4 & & \end{tabular}\\ 
\hline
\hline
\end{tabular}
\end{table*}
\normalsize

\normalsize

\begin{enumerate}

\item directly, considering the angle-energy integrated cross section for the detected projectile charged fragments, obtained in the following Refs.\cite{67Li208Pb, 7Li124Sn,6Li59Co, ribras3, 6He64Zn, 6He64ZnAguilera,8Li209Bi, osvaldo, KOL02,8B58Ni, 8B64Zn, 8He197Au,Jin15,6He209BiKolata,kuma,11Li9Be12C,11Be64Zn,11Li208Pb,7Be28Si,8B28Si,17F58Ni,alfa64ZnMet1}. In the $^9$Be$+^9$Be case Ref. \cite{9Be9Be},  a detailed $\gamma$-spectroscopy analysis was performed to identify all the relevant reaction channels. 
\item indirectly, from the difference, $\sigma_{dir}=\sigma_ {R}-\sigma_{fus}$,  for the cases where total reaction and complete fusion cross sections (CF) are available \cite{6He209Bi,4He209Bi_Elastic,6He209Bi,verma_interaction_2010,67Li208Pb, 16ONi,12C16O,12C208Pb,12C209Bi,8B58Ni,8He208Pb,9Be208Pb_Fusao,9Be208Pb_FusaoII,12C13CMet2}. In the present study, we considered only CF cross sections.
\end{enumerate}
%46He188Os, 9Be27Al,

\setlength{\tabcolsep}{3pt}

\begin{table*}[!ht]
\small
    \centering
    \begin{tabular}{lp{1.0cm}p{13.0cm}}
    \hline
    \hline
    System  & Ref. & Note\\
      \hline
      \multicolumn{3}{c}{Type: exotic \textit{n}-rich}\\
      \hline
      $^6$He + $^{209}$Bi & \cite{6He209Bi} & The raw data is total fusion cross sections evaluated at Ref.\cite{6He209BiKolata} by fusion products dis\-cri\-mi\-na\-ted on the basis of delayed $\alpha$ yields. ICF (Incomplete Fusion) of the $^4$He core was not considered due to the Q-value for this reaction channel. \\
      \hline
      %%^{8}$He + $^{208}$Pb  & \cite{8He208Pb} & Fusion cross sections were evaluated at Ref.\cite{8He208Pb} by the discrimination of the most probable fusion-evaporation channel $^{8}$He($^{208}$Pb, 4n)$^{212}$Po.\\
      %%\hline
      \multicolumn{3}{c}{Type: WB}\\
      \hline
     $^6$Li + $^{208}$Pb & \cite{6He209Bi} & Fusion cross sections were evaluated at Ref.\cite{6Li_Fusion} by fusion residuals yields. ICF residuals were not observed. \\
      \hline
     $^7$Li + $^{208}$Pb & \cite{67Li208Pb} & Barrier-penetration model calculations which used its own optical model parameters. This method can show reasonable accuracy \cite{WKB}.\\
      \hline
     $^7$Li + $^{9}$Be & \cite{verma_interaction_2010} & Fusion cross-sections are evaluated by $\alpha$ yields at backwards angles and PACE calculations. \\
     \hline
     $^{9}$Be + $^{208}$Pb & \cite{9Be208Pb_Fusao, 9Be208Pb_FusaoII} & CF cross sections were obtained by summing the cross sections for Rn isotopes residues originated from important neutron evaporation channels and the fission one. \\
    \hline
    \multicolumn{3}{c}{Type: exotic \textit{p}-rich}\\
      \hline
     $^8$B + $^{58}$Ni & \cite{8B58Ni} & Proton, $\alpha$ yields and PACE calculations were used to evaluate CF fusion cross sections. ICF and CF were discriminated by the proton evaporation.\\
    \hline
    \multicolumn{3}{c}{Type: SB}\\
      \hline
     $^{16}$O+$^{58}$Ni & \cite{16ONi} & Barrier penetration model using the best fit optical model potentials. Recent experimental fusion data at Ref.\cite{16O58Ni_Recente_REFEREE} are available for beam energies below those of Ref.\cite{16ONi}, where BPM calculations are adhent to these experimental data.\\
     \hline
    $^{12}$C+$^{209}$Bi & \cite{12C209Bi} & Evaporation residues and fi\-ssion frag\-ments mea\-su\-re\-ments.  ICF is not considered a issue in this system.\\
    \hline
    $^{12}$C+$^{13}$C & \cite{12C13CMet2} & Total fusion cross sections evaluated considering the integrated angular distributions of evaporation residues with $\mathrm{Z} \ge 6$.  \\
     \hline
    $^{12}$C+$^{16}$O & \cite{12C16OFusion} & Fusion fragments measured  using energy and time-of-flight tech\-ni\-ques. ICF is not considered a issue in this system.\\
    \hline
    \hline
    \end{tabular}
    \caption{Summary of CF/ICF data (Complete Fusion/Incomplete Fusion). In the present study we are only interested in CF cross sections.}
    \label{tab:Fusion}
\end{table*}

\begin{figure*}[!ht]
    \centering
    \includegraphics[width=0.7\textwidth]{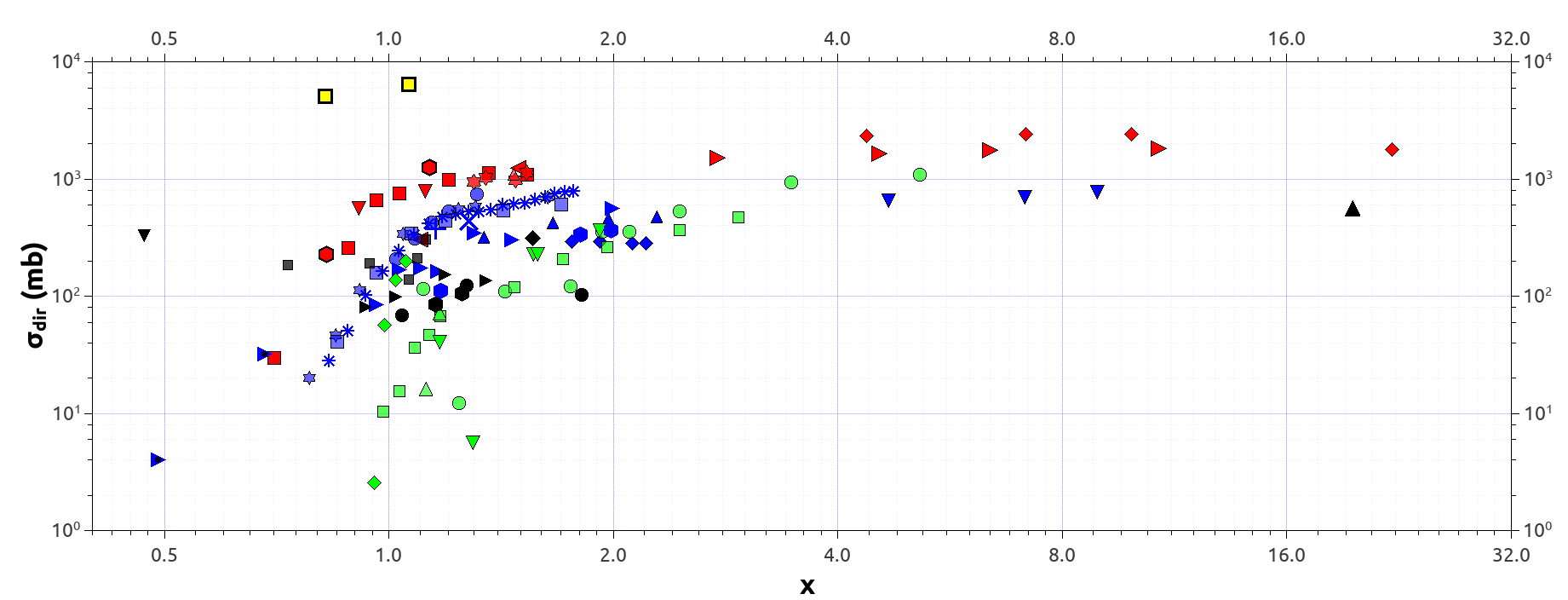}
    \caption{Direct cross section data (mb) as function of $x=E/V_b$. Different types of projectiles are shown by different colors, exotic \textit{n}-rich (red), weakly bound (WB) (blue), strongly bound (SB) (green) and exotic \textit{p}-rich (black). The symbols are the same as in table \ref{tab:TodosDados}. The yellow squares represent the $^{11}$Li+$^{208}$Pb system - Ref.\cite{11Li208Pb} - see text for details. The dotted magenta line means $x=1.8$.}
    \label{fig:Secoes_de_Choque}
\end{figure*}

In Table \ref{tab:TodosDados}, we present a list of all systems considered 
and in Fig.\ref{fig:Secoes_de_Choque}, a plot of the selected experimental data is presented as a function of the variable $x=E/V_b$. In Fig. \ref{fig:Secoes_de_Choque}, we see that the data are scattered; however, it is possible to observe four different data groups of cross sections: the exotic \textit{n}-rich projectiles (red) with the largest cross sections except for the $^{11}$Li case (yellow squares) which will be discussed latter; the weakly bound (WB - blue) projectiles present a considerably smaller cross section; and, finally,  the strongly bound (SB - green)  with the smallest cross sections. There are also very few points for the proton rich $^8$B and $^{17}$F projectiles (\textit{p}-rich black),
with cross sections comparable to the strongly bound. This result is quite surprising considering that these proton rich projectiles are very loosely bound and a large direct cross section for breakup and other reaction channels is expected above the Coulomb barrier. However, there are very few points above the barrier which prevents any de\-fi\-ni\-te conclusion at this point.
%In the case of $^{17}$F, it was considered that its first excited state, which is considered a exotic \textit{p}-rich, plays a major role in direct channels and, for this reason, $^{17}$F was classified as a exotic \textit{p}-rich in present work.

\begin{figure*}[!ht]
    \centering
    \includegraphics[width=0.7\textwidth]{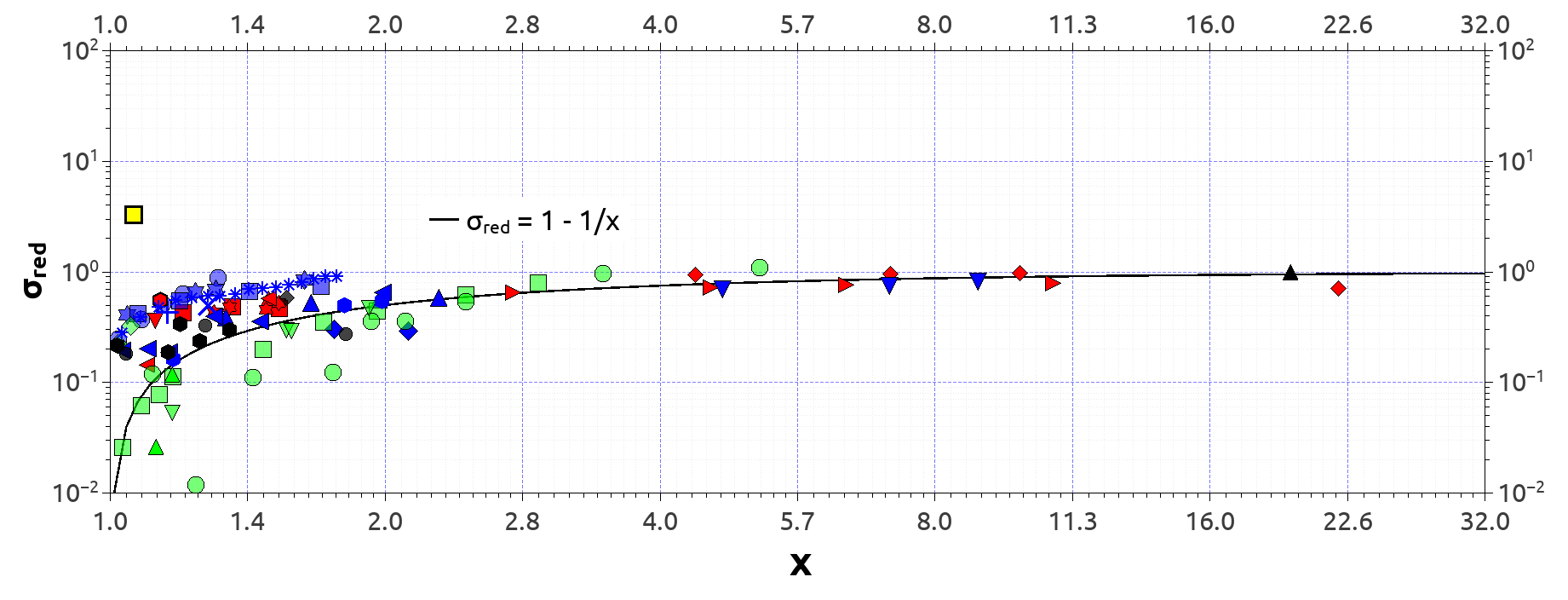}
    \caption{Reduced direct cross sections (dimensionless) as a function of the dimensionless reduced energy - $x=E/V_b$. The same symbols and color system applied at table \ref{tab:TodosDados} and Fig.\ref{fig:Secoes_de_Choque} are used. The black continuous line is the function $\sigma_{red} = 1-1/x$. In this plot $\hat{a}_n^{n-halo} = 1.32(06)$ fm; $\hat{a}_n^{WB} = 0.66^{+0.16}_{-0.12}$ fm; $\hat{a}_n^{SB} = 0.68^{+0.16}_{-0.04}$ fm; $\hat{a}_n^{p-halo} = 0.55$ fm. As in Fig.\ref{fig:Secoes_de_Choque}, this figure uses, in its vertical scale, four orders of magnitude. The universal curve for the reduced direct cross sections is meaningless for $x<1$.} 
    \label{fig:Final}
\end{figure*}

To compare the data with our model, all cross sections from Fig.\ref{fig:Secoes_de_Choque} were divided by $\sigma_{dir}=2\pi R_xa$. $R_x$ and $a$ were calculated using Eqs.[\ref{eq5a}, \ref{eq6a}, \ref{eq7a}] with a single parameter $a_n$, the nuclear diffuseness, adjusted to best fit the universal curve Eq.[\ref{uni1}].
The results are shown in Fig. \ref{fig:Final}. As one can see by comparing Figs.\ref{fig:Secoes_de_Choque} and \ref{fig:Final}, the division by $\sigma_{dir}=2\pi R_xa$ clearly condenses the data in well defined groups.  In fact, the bi-dimensional reduced variance \cite{BidimVar} between the $(x,\sigma_{dir}/\sigma_{dir,max})$ and $(x,\sigma_{dir}/\sigma_{red})$ is decreased
 by a factor of approximately 5 in the $x>1.8$ range; however, in the $x$ lower range, there is not any significant change the bi-dimensional reduced variance.   

For energies well above the barrier, $x\ge 1.8$, all the points fit the universal curve [\ref{uni1}],  using different values of $a_n$ for each type of projectile, exotic \textit{n}-rich, stable weakly bound (WB), stable strongly bound (SB) and exotic \textit{p}-rich.  The following results were obtained: $\hat{a}_n^{n-rich} = 1.32(06)$ fm; $\hat{a}_n^{WB} = 0.66^{+0.16}_{-0.12}$ fm and $\hat{a}_n^{SB} = 0.68^{+0.16}_{-0.04}$ fm.
The above values of $a_n$ and errors were obtained by minimizing the chi-square, separately for each type of projectile. In the \textit{p}-rich case, only two points are above $x \ge 1.8$ which were adjusted to the universal curve to obtain $a_n^{p-rich}=0.55\,$fm, with no error estimation. This $a_n^{p-rich}$ leads to an estimation of $R_x = 7.16\,$fm and $a = 1.23\,$fm for the $^{8}$B + $^{28}$Si system that is reasonable close to the ones evaluated at Ref.\cite{8B58Ni} where the values of 7.5 and 1.02$\,$fm were obtained, respectively.

%It is interesting to note that the behavior seen in Fig. \ref{fig:Secoes_de_Choque}, where the neutron rich projectiles present the larger cross sections, is partially accounted ($x>1.8$) in our model by the dependence on the parameter $X$, which increases from proton rich to neutron rich systems (see Eq.\ref{eq6a}).

%\textcolor{red}
%{During the fitting process, it was not possible to e\-va\-lua\-te the individuals reduced cross-sections uncertainties, because some systems does not have their uncertainties for both reaction and/or fusion cross sections or for each direct reaction channels. For this reason, each data point in this work has the same weighting factor.}

The results show that a  considerably larger value of $a_n$ is necessary to fit the exotic \textit{n}-rich projectiles to the universal curve, in comparison to the weakly bound, strongly bound and exotic \textit{p}-rich. Those three last cases gave values of diffuseness near the expected standard value of $a_n\approx 0.65$ fm. \textit{P}-rich exotic projectiles namely $^8$B and $^{17}$F present the lowest diffuseness, similar to the strongly bound stable systems. This result contrasts with previous analyses where the total reaction cross section for exotic \textit{p}-rich projectiles presents a large value compared to the stable systems \cite{vivi2,uiran2022x}. However, no conclusion is possible here due to the small number of experimental points considered. 

On the other hand, in the region $x<1.8$ the situation seems different. A considerable amount of data for exotic \textit{n}-rich and WB projectiles present si\-mi\-lar reduced cross sections, both considerably above the universal curve and this behavior persists for energies below the Coulomb barrier. We see two possible explanations for this behavior. Coupled channels effects  are expected to be more important at lower energies and could cause this enhancement. Another possible explanation could be the fusion contamination in the direct cross section data, as the experimental separation between these processes becomes more difficult at lower energies. %Moreover, the Coulomb barrier evaluation is a systematic procedure which could introduce bias in exotic systems. 

\begin{figure}[!ht]
    \centering
    \includegraphics[width=0.5\textwidth]{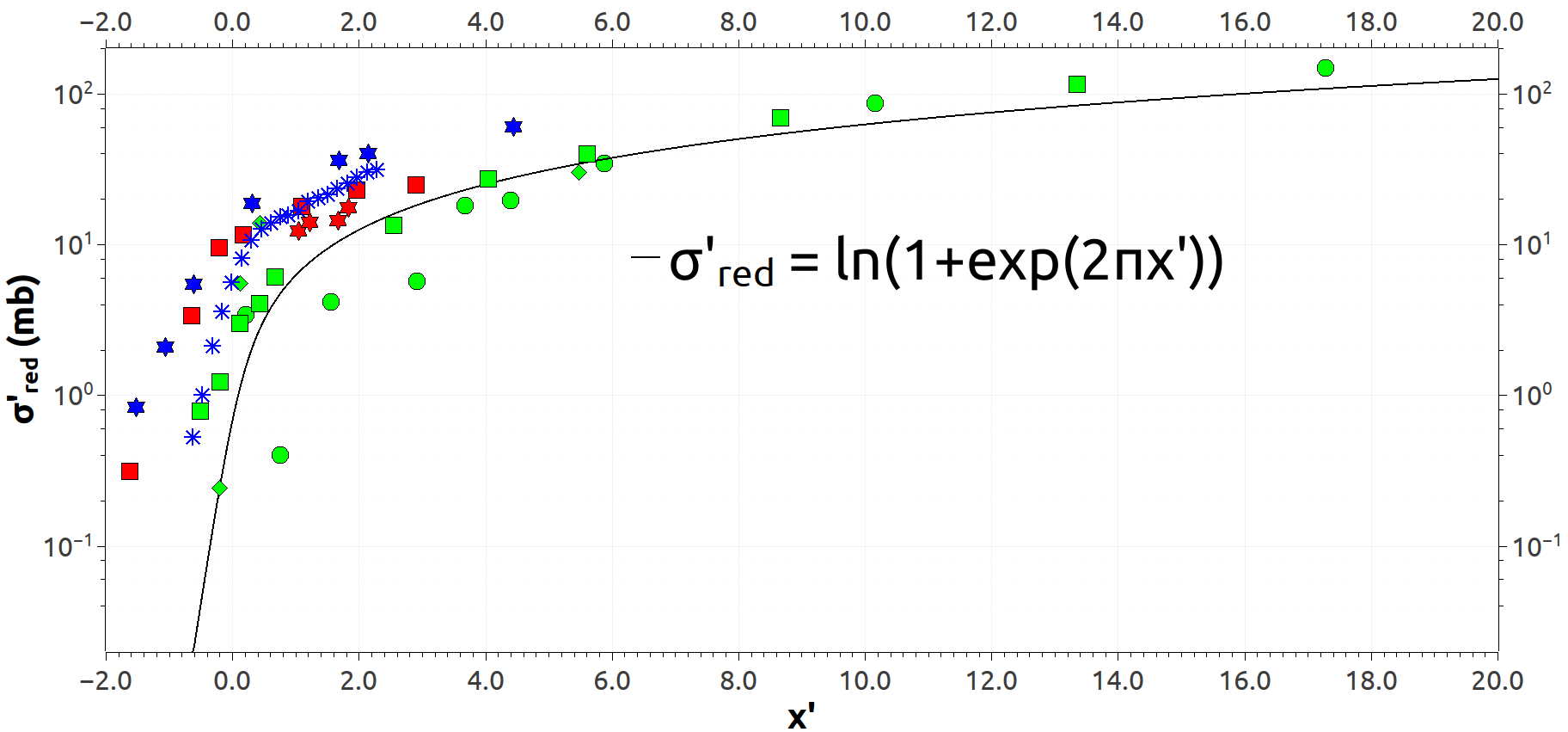}
    \caption{Reduced cross sections $\sigma'_{red}$ as a function of the reduced energy - $x'=(E-V_b)/\hbar\omega$ using the Wong's formula. The same symbols and color system applied at table \ref{tab:TodosDados} and Fig.\ref{fig:Secoes_de_Choque} are used. The black continuous line is the function $\sigma'_{red} = \mathrm{ln}(1+\mathrm{exp}(2\pi x'))$. See text for more details.} 
    \label{fig:Wong}
\end{figure}

One may argue that, for the region below the Coulomb barrier, the $(x',\sigma'_{red})$ reduction scheme of Eq.[\ref{uni2}] using the Wong formula would be more appropriate. The result using the Wong formula is presented in Fig.\ref{fig:Wong}, where the same $a_n$ va\-lu\-es fitted at high energies were used. We see that the enhancement is not explained also by the Wong formula.

Finally, the $^{11}$Li$+^{208}$Pb case is remarkable as it falls much above the universal curve. The data for the $^{11}$Li$\rightarrow$ ${^9}$Li$+ n+n$  breakup reaction on a $^{208}$Pb target was obtained in a nice clean experiment and are presented in Ref.\cite{11Li208Pb}. There is no doubt about those experimental cross sections, and the results show how large the $^{11}$Li  breakup cross section on heavy targets can be, possibly, due to the contribution of the breakup in the Coulomb field of the heavy target. The measured $^{11}$Li breakup cross section of $\approx 5$ barns probably exhausts the total reaction cross section in this case.
In the case of $^{11}$Li there are also measurements on light targets $^{12}$C and $^9$Be \cite{11Li9Be12C}
above the barrier which fit very well with $\hat{a}_n^{n-rich} = 1.32(06)$ fm as obtained for \textit{n}-rich projectiles. This indicates that the enormous effect seen in the $^{11}$Li$+^{208}$Pb case is probably caused by the Coulomb breakup contribution. It is interesting to mention that,  for the $^7$Li projectile, the situation is different and both $^7$Li$+^{208}$Pb and $ ^7$Li$+^9$Be cases gave similar reduced cross sections in better agreement with the universal curve. 

%\clearpage
\section{Conclusions}
\label{sec:Conclusao}
A model to estimate  the contribution of direct  reactions to the total reaction cross section of different projectile-target systems is proposed. The method separates the total reaction cross section into two contributions, one from the total fusion cross section, which scales as the disk area $\pi R^2$ and another from the direct processes, which scale as the area of a ring $2 \pi Ra$, $a$ being the width of the ring in \textit{r}-space. The method is applied for reactions induced by stable and exotic projectiles on different mass targets and e\-ner\-gies around and above the Coulomb barrier. It was found that the direct part of the cross section scales well with the $2 \pi Ra$ expression for energies above the Coulomb barrier, and for a large range of target masses. The width parameter $a$ is directly related to the nuclear diffuseness $a_n$ and seems to be dependent basically on the projectile structure. A considerably larger  nuclear diffuseness $a_n$ was obtained for the exotic \textit{n}-rich  $^{6,8}$He, $^{11}$Be and $^{11}$Li projectiles compared to the stable weakly bound and strongly bound projectiles such as $^4$He, $^{12}$C and $^{16}$O. 

Our results indicate that, well above the Coulomb barrier, the enhancement observed in the direct cross section for the neutron halo projectiles is considerable and can be accounted by a larger nuclear diffuseness $a_n$ parameter in the model. 

On the other hand, for $x < 1.8$, the model does not account for the observed enhancements in neutron halo and weakly bound systems. We believe that, for energies below the Coulomb barrier, coupled channel effects become more important and should be explicitly taken into account in order to reproduce the data, which is not the case for the present geometric model. Moreover, at lower energies, the experimental separation between direct and non-direct processes becomes more difficult and some contamination from fusion in the direct cross section could also explain a part of this enhancement.  

The geometric model presented here is not intended to reproduce all the complexity and details of specific direct reaction channels. However, it may be useful to scale the contribution of total direct processes to the reaction cross section, allowing the comparison of different systems with a common framework.

%\clearpage
%------------------------------
\section*{Acknowledgments}
%----------------------
%This work has been partially supported by Conselho Nacional de Desenvolvimento Cient\'{\i}fico e Tecnol\'ogico -- CNPq/MCTI (Brazil), Funda\c{c}\~ao de Amparo \`a  Pesquisa do Estado de São Paulo, FAPESP (Brazil), contracts no. 2019/02759-0, 2019/07767-1, 2016/17612-7, 2013/22100-7,  Coordenação de Aperfeiçoamento de Pessoal de Nível Superior – Brasil (CAPES) – Finance Code 88887.355019/2019, and Funda\c{c}\~ao de Amparo \`a  Pesquisa do Estado do Rio de Janeiro, FAPERJ (Brazil).

This work has been partially supported by Funda\c{c}\~ao de Amparo \`a Pesquisa do Estado de S\~ao Paulo, Brazil (FAPESP) - contracts no. 2019/07767-1, 2019/02759-0, and 2021/12254-3; Coordena\c{c}\~ao de Aperfei\c{c}oamento de Pessoal de N\'ivel Superior, Brazil (CAPES) – Finance Code 88887.355019/2019 and 88887.620098/2021 and Conselho Nacional de Desenvolvimento Cient\'ifico e Tecnol\'ogico, Brazil (CNPq) and by the project INCT-FNA Proc. No. 464898/2014-5. We would like to thank Prof. V. V. Parkar for providing us with experimental data for the  $^7$Li + $^{124}$Sm system and Prof. Wayne Allan Seale for the text review. 

\clearpage
%\nocite{*}
\bibliographystyle{elsarticle-num}
\bibliography{refer1}% Produces the bibliography via BibTeX.

\end{document}